\documentclass{article}
\usepackage{epsfig}
\usepackage{amsmath,amssymb}
\usepackage{cite}
\usepackage{bm}

\begin{document}
	
	\begin{center}
		{\bfseries The neutron radiation capture process $n+p\rightarrow D+\gamma$
			and $^1 S_0$ $np$ wave function}
		
		N.A. Khokhlov
		
		{\small  {\it Komsomolsk-on-Amur State Technical University, Russia. }
			\\
			{\it E-mail: nikolakhokhlov@yandex.ru }}
	\end{center}
	
	\begin{abstract}
	Cross section of the neutron capture reaction $n+p\rightarrow d+\gamma$ at threshold was calculated with different realistic
	$^{3}S_{1}$ deuteron and $^{1}S_{0}$ $np$ scattering wave functions stemming from  Nijmegen-II, JISP16, Paris,  Idaho and
	Moscow (with forbidden states) potentials. It is found that this reaction with thermal neutrons may be described  without a contribution of meson exchange currents.
	\\[\baselineskip]
\end{abstract}

\thispagestyle{plain}

\section{\label{sec:intro}Introduction}
 There is a well known assumption that
thermal $np$ capture and deuteron electrodisintegration near threshold
 provide convincing evidence for the pion exchange
currents in the two nucleon electromagnetic reactions \cite{Mathiot1984,PionsAndNuclei}. Nobody doubts that there is some contribution of these currents into such reactions.
But it is not clear enough what is their exact contribution.

For example various realistic $NN$-potentials (Nijmegen-I (NijmI), Nijmegen-I (NijmII) \cite{Nijm}, JISP16 \cite{JISP16}, CD-Bonn \cite{CD-Bonn}, Paris \cite{Paris},
Argonne18 \cite{Argonne18,Argonne18ms}, Idaho \cite{Idaho}
and Moscow (with forbidden states) \cite{NNscatMy3}) without such currents give quite distinguished results for the
 deuteron electromagnetic form factors  above 5~$Fm^{-1}$ of the four-momentum transferred by a probe ($q^2=-Q^2$) \cite{My2016}.
 This implies that the mentioned potentials require quite different meson exchange currents. The NN-potential and exchange currents can not be extracted directly from
 experiments. From the microscopic QCD point of view these notions are effective and useful simplification tools.
 One can see  \cite{NNReview2014} that there are models with different  non-nucleon degrees of freedom or at least with
 different their contribution into the $NN$-interaction. It seems that simplicity of the model may be more important than their shaky microscopic justification.
 The separable JISP16 potential is perhaps the most prominent examples of such an opportunist approach.
  
 Local $NN$ potentials seem to be the simplest ones and they may describe the $NN$ elastic scattering by construction up to 2.7~GeV of the laboratory energy  at least \cite{NNscatMy3}.
 But as mentioned earlier the general agreement is that without exchange currents contribution they fail  in description of  simple electromagnetic reactions with two nucleons.
 In this paper we dispute this point of view and show that the part played by the meson exchange currents in the low energy
 $n+p \rightleftarrows d+\gamma$ reactions may be simulated by properly fitted local $^{3}SD_{1}$ and $^{1}S_{0}$ $np$ partial potentials that describe also the $NN$ elastic scattering by construction
 and deuteron electromagnetic form factors.
 
 The cross section for the $n+p \rightarrow d+\gamma$ reaction at the threshold is written \cite{Mathiot1984,PionsAndNuclei} as
 \begin{equation}
 \sigma(np\rightarrow d\gamma)=\frac{2\pi \alpha w^3}{M_{N}q}\left| F \right|^2 ,
 \end{equation}
 where
 \begin{equation}
  F = \frac{\mu_{p}-\mu_{n}}{2}\int_{0}^{\infty} dr\,u_{0}(r)u(r).
 \end{equation}
 In these expressions $w$ and $q$ are the photon energy and nucleon momentum in the center of masses system. $M_{N}$ is a nucleon mass, $\alpha$ is a fine structure constant.
 Radial $^{1}S_{0}$  wave function $u_{0}(r)$  is normalized as
 \begin{equation}
 u_{0}(r)\rightarrow \frac{1}{q}\sin(qr+\delta_{0}).
 \end{equation}
 Radial deuteron $^{3}S_{1}$ wave function  $u(r)$  is normalized as
  \begin{equation}
  u^{2}(r)+w^{2}(r)=1,
  \end{equation}
 where  $w(r)$ is a radial deuteron $^{3}D_{1}$ wave function. We use natural units, where $\hbar = c =1$.
 A nucleon momentum $q\approx 1.745 \cdot 10^{-5}\ Fm^{-1}$ for thermal neutrons.

\begin{figure}[htb]
 	\centerline{\includegraphics[width=1.0\textwidth]{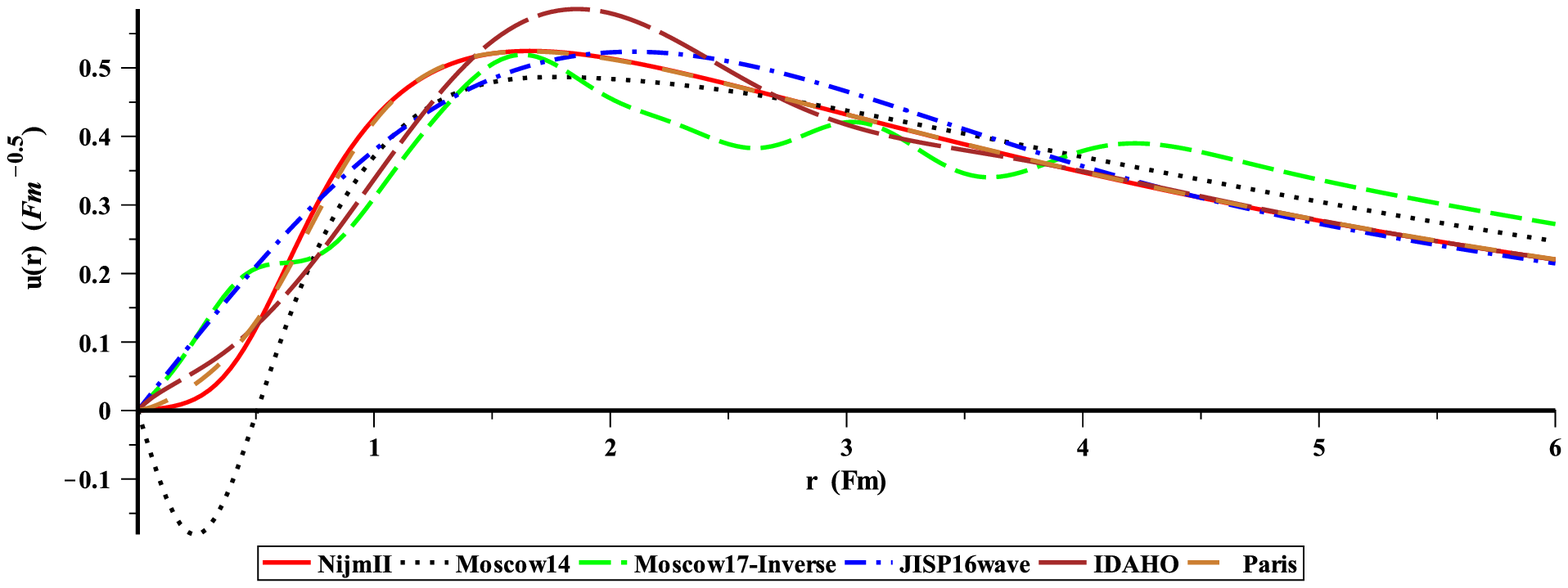}}
 	\centerline{\includegraphics[width=1.0\textwidth]{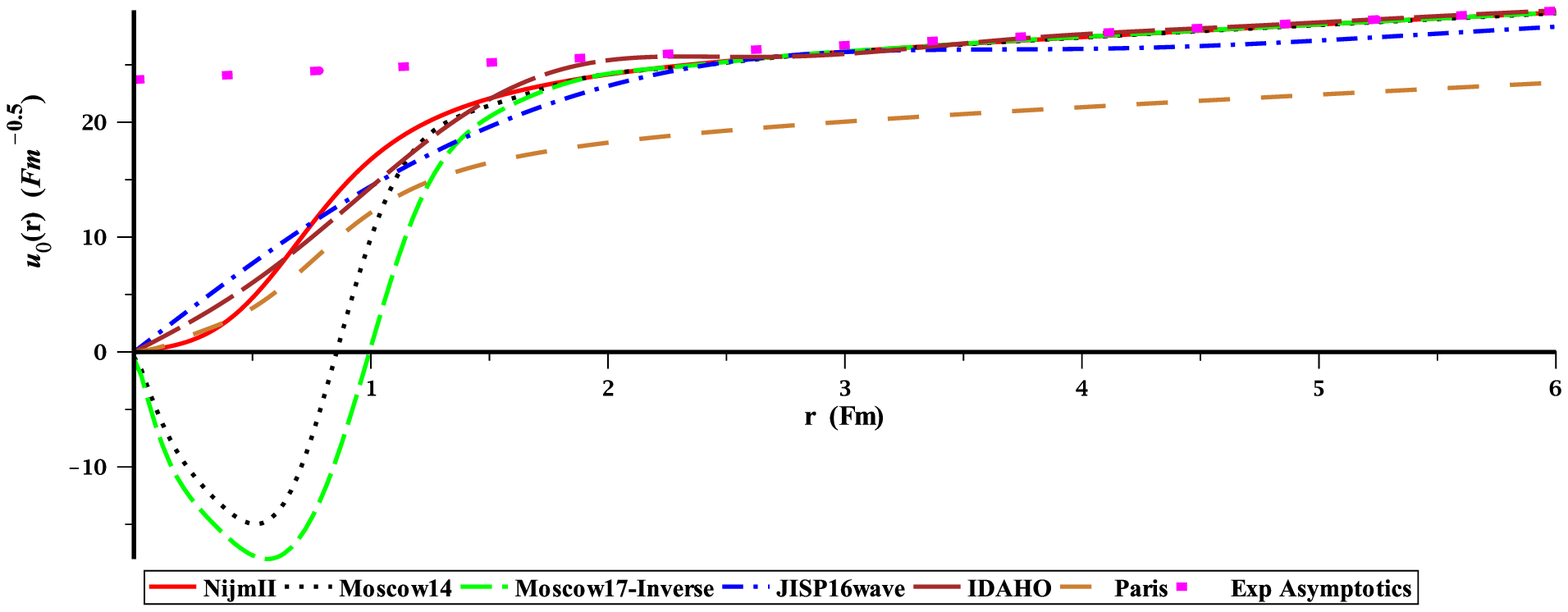}}
 	\caption{\label{fig:nFFs} $^{3}S_{1}$ deuteron $u(r)$ and $^{1}S_{0}$ $NN$ scattering partial $u_{0}(r)$ wave functions stemming from realistic $NN$ potentials used in the calculation. "Exp Asymptotics" curve is defined as $r-a_{s}$, where $a_{s}\approx 23.7\ Fm$ is the $np$ singlet scattering length.}
 \end{figure}

\begin{figure}[htb]
 	\centerline{\includegraphics[width=1.0\textwidth]{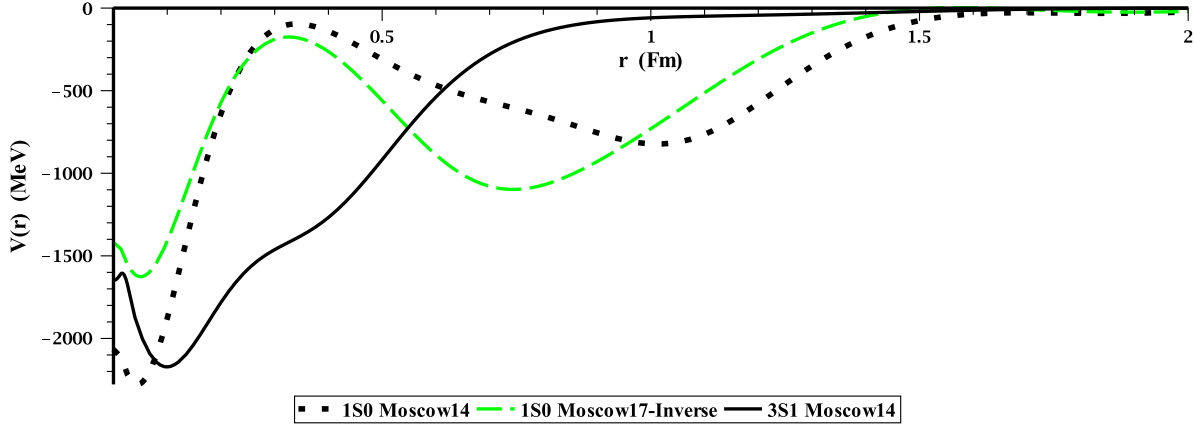}}
 	\caption{\label{fig:potentials} Moscow potentials.Two $^{1}S_{0}$ partial potentials are exactly phase equivalent.}
 	\end{figure}

 The  wave functions used in the calculation are shown in Fig.~\ref{fig:nFFs}. We see that wave functions are quite different for all
 the potentials not only in the inner region ($r<4\ Fm$) but also in the asymptotic region.
 The charge independent Paris potential  \cite{Paris} gives $nn$ singlet scattering length ($T=1$).  
 One must note that this Paris potential was used in  Ref.~\cite{Mathiot1984} to establish the necessity of taking into account meson exchange currents for description of the deuteron electrodisintegration near threshold. JISP16 is also failing in description of the experimental value of the $np$  scattering length  $a_{s}\approx 23.7\ Fm$. Wave functions stemming from NijmII and Idaho potentials differ only in the inner region. There are two sets of Moscow type wave functions. One of them labeled Moscow14  is stemming from a charge-dependent version of the  Moscow potential of Ref.~\cite{NNscatMy3} with slightly changed $^{3}SD_{1}$ partial potentials (fitted to $np$ triplet scattering length and to exact values of all three static deuteron electromagnetic form factors) and changed $^{1}S_{0}$ partial potential (fitted to experimental $np$ singlet scattering length and to the thermal $np$ capture).  Another set labeled as Moscow17-Inverse is the
 $^{3}S_{1}$ deuteron wave function extracted earlier  from the eD elastic scattering data \cite{My2016} and $^{1}S_{0}$ scattering wave function stemming from the partial potential fitted to the  thermal $np$ capture. Corresponding partial potentials are presented in Fig.~\ref{fig:potentials}. Both Moscow potentials describe considered experimental data by construction.
 Values were fitted in the inversion procedure of Ref.~\cite{NNscatMy3} for extraction of the $^{1} S_{0}$ partial potentials from $NN$ scattering data.
 A user-friendly computer code for calculation of the potentials is
 available from the author upon request.
 	\begin{table}[ht]
 		\caption{Cross sections of the neutron capture process $n+p\rightarrow D+\gamma$ calculated with different realistic deuteron and $NN$ scattering wave functions.}
 		\label{tab1}
 		\begin{center}
 			\begin{tabular}{|c|c|c|}
 				\hline\noalign{\smallskip}
 				&$a_{s}$, Fm &$\sigma(np\rightarrow D\gamma$, mb  \\
 				\noalign{\smallskip}\hline\noalign{\smallskip}
 				Exp &  -23.7 & $334.2\pm 0.5$ \\
 				NijmII & -23.6   & 304 \\
 				Idaho & -23.7    & 308  \\
 				Paris & -17.5  & 188 \\
 				JISP16 & -22.4 & 286 \\
 				Moscow14 &-23.7  & 334 \\
 				Moscow17-eD Inverse & -23.7  & 334 \\
 				\noalign{\smallskip}\hline
 			\end{tabular}
 		\end{center}
 	\end{table}
Results of our calculations shown in Table~1. We see from our results that it is quite possible to describe the low energy
 $n+p \rightarrow d+\gamma$ reaction (and therefore inverse process) by a properly fitted local $^{1}S_{0}$ $np$ partial potential that describes also the $NN$ elastic scattering up to 2.7 GeV by construction (Moscow14 potential Figure \ref{fig:potentials}).
Moreover it seems possible to describe also all available eD elastic scattering data with fitted realistic $^{3}SD_{1}$ deuteron waves and with $^{1}S_{0}$ wave stemming from an $np$ local potential and fitted simultaneously to the $NN$ elastic scattering up to 2.7 GeV (Fig.~\ref{fig:potentials}) and to $n+p \rightleftarrows d+\gamma$ data (Moscow17-eD Inverse potential). Unfortunately, the Moscow17-eD Inverse $^{3}SD_{1}$ potentials are not constructed yet in a local form. We have only its $^{3}SD_{1}$ deuteron waves \cite{My2016}. The work is in progress. 
It is interesting that deuteron partial wave functions ("$eD$ inverse") extracted from the $eD$ elastic scattering \cite{My2016} 
differ from  the Moscow type wave functions (for example of Ref.~\cite{NNscatMy3}) in position of the node. 
The node of the $eD$-inverse deuteron wave functions is in the $D$-wave, whereas it is in the $S$-wave for usual concept of the Moscow type potential. We suppose that this relocation may be brought on by a tensor interaction or other effects, though its possible microscopic origin is unclear.   
 
Another interesting problem is the threshold electrodisintegration of the deuteron  that also
''gives particularly strong evidence for the existence of pion exchange currents'' \cite{PionsAndNuclei}. The situation here is not as simple as for the $n+p \rightleftarrows d+\gamma$ reactions and 
not as simple as it is stated in \cite{Mathiot1984,PionsAndNuclei}. It is shown that relativistic \cite{Arriaga2007} and other \cite{Leidemann1983} effects are not negligible for all the experimentally investigated region. These effects completely change results of  \cite{Mathiot1984}.

\bibliographystyle{unsrt}
\bibliography{khokhlov.bib}{}
\end{document}